Article

# Experimentally informed structure optimization of amorphous TiO$_2$ films grown by atomic layer deposition

Jun Meng[a‡*], Mehrdad Abbasi[b‡], Yutao Dong[a], Corey Carlos[a], Xudong Wang[a], Jinwoo Hwang[b*], Dane Morgan[a*]

Amorphous titanium dioxide TiO$_2$ (a-TiO$_2$) has been widely studied, particularly as a protective coating layer on semiconductors to prevent corrosion and promote electron-hole conduction in photoelectrochemical reactions. The stability and longevity of a-TiO$_2$ is strongly affected by the thickness and structural heterogeneity, implying that understanding the structure properties of a-TiO$_2$ is crucial for improving the performance. This study characterized the structural and electronic properties of a-TiO$_2$ thin films (~17nm) grown on Si by Atomic Layer Deposition (ALD). Fluctuation spectra $V(k)$ and angular correlation functions were determined with 4-dimensional scanning transmission electron microscopy (4D-STEM), which revealed the distinctive medium-range ordering in the a-TiO$_2$ film. A realistic atomic model of a-TiO$_2$ was established guided by the medium-range ordering and the previously reported short-range ordering of a-TiO$_2$ film, as well as the interatomic potential. The structure was optimized by the StructOpt code using a genetic algorithm that simultaneously minimizes energy and maximizes match to experimental short- and medium-range ordering. The StructOpt a-TiO$_2$ model presents an improved agreements with the medium-range ordering and the $k$-space location of the dominant 2-fold angular correlations compared with a traditional melt-quenched model. The electronic structure of the StructOpt a-TiO$_2$ model was studied by ab initio calculation and compared to the crystalline phases and experimental results. This work uncovered the medium-range ordering in a-TiO$_2$ thin film and provided a realistic a-TiO$_2$ structure model for further investigation of structure-property relationships and materials design. In addition, the improved multi-objective optimization package StructOpt was provided for structure determination of complex materials guided by experiments and simulations.

## 1. Introduction

As a promising path to generate sustainable fuels, photoelectrochemical (PEC) systems demand efficient and low-cost photoelectrode materials to achieve commercialization.[1] Silicon is a feasible semiconductor for PEC water splitting due to its narrow band gap for efficient absorption of visible light and industrial maturity.[2] Nonetheless, Si commonly encounters deleterious surface corrosion during the electrochemical water splitting reaction, forming an insulation layer of silicon dioxide. This surface corrosion constrains the operational lifetime and efficiency of PEC devices[3] and there is therefore interest in creating protective coatings to slow or stop the corrosion. In recent years, amorphous oxide thin films, synthesized by atomic layer deposition (ALD), have been used as a critical component in many modern energy devices.[4–6] In particular, amorphous TiO$_2$ (a-TiO$_2$) thin films emerged as a superb protection layer to separate the photoelectrode surface from the corrosive electrolyte. It has been proven that the a-TiO$_2$ coated photoelectrode can dramatically improve the electrochemical efficiency and corrosion resistance by facilitating interfacial charge/mass transfer and stabilizing the photoelectrode surfaces of the a-TiO$_2$ layer.[7–13] These amorphous ultrathin films have usually been considered to be homogeneous in terms of their structure, and have uniform properties throughout the film.[8] However, studies have shown that the property of those amorphous thin films can be changed substantially by its synthesis conditions such as temperature, cycles, reductive or oxidative conditions,[14,15] thickness and crystallinity,[16–18] and reaction conditions.[19] For example, a homogeneous 2.5 nm a-TiO$_2$ thin film achieved much better electrode longevity than a 24 nm thick a-TiO$_2$ film.[17] The thicker (24 nm) films contained metastable intermediate phases, which indicates that the local structural heterogeneity involving metastable phases could result in different electronic properties.[17] Such studies demonstrate the importance of understanding the detailed local structure and heterogeneity.

Understanding the detailed atomic structure of amorphous materials, however, remains challenging for both experimental and computational approaches. Experimentally, the structure of amorphous materials has been extensively studied using the pair distribution function (PDF). PDFs give the probability of finding another atom at a certain distance away from the origin atom and are typically measured from large area diffraction experiments using X-rays or neutrons.[20–24] However, the information from PDFs is limited to the atomic bonding at short distances (i.e. ordering within a few neighboring shells), or short-range ordering (SRO). Due to the nature of inherent averaging in PDF, it is difficult to study the structural features beyond the averaged SRO, including structural ordering at the nanometer scale.[25–27] Such nanometer scale ordering can involve local nanoscale volumes with a relatively high degree or structural (or even chemical) ordering, which is commonly known as medium-range ordering (MRO). Characterization of

a. Department of Materials Science and Engineering, University of Wisconsin-Madison, Madison, WI 53706, United States.
b. Department of Materials Science and Engineering, The Ohio State University, Columbus, Ohio 43210, United States
‡ These authors contribute equally to this work.

MRO requires nanometer-sized experimental probes, which can be obtained by combining electron nanodiffraction with related analysis methods. The fluctuation electron microscopy (FEM) method measures the magnitude of intensity fluctuation related to the size and distribution of MRO[25] and the angular correlation analysis reveals the dominant average symmetry of MRO.[26–28] These methods based on nanodiffraction have demonstrated the importance of examining nanoscale heterogeneity and MRO in order to properly understand the properties of amorphous materials. Previous experimental studies of the a-TiO$_2$ were mostly focused on the SRO[29,30] and no MRO information was directly studied.

a-TiO$_2$ structure has also been investigated by different computational approaches such as density functional theory (DFT) based Ab Initio Molecular Dynamics (AIMD),[31–33] classical interatomic potential molecular dynamics (IPMD), and Reverse Monte Carlo (RMC) methods.[29,30] However, all of these approaches have significant limitations. Due to the computational limitations of the simulation time and size, AIMD simulated amorphous structures require extremely fast quenching rates (approximately $10^{12}$ deg/s or faster) and small unit cells (approximately a couple of hundred atoms or fewer), which can potentially significantly impact the obtained amorphous structure, particular MRO which might require larger cells and more time to form. IPMD can reach longer length and time scales, and the length scales are likely large enough for realistic results while the cooling rates are still much faster than almost all experiments (approximately $10^9$ deg/s or faster). Furthermore, IPMD relies on an empirical interatomic potential, and it is difficult to be sure such potentials are accurate for modeling amorphous phases. AIMD and IPMD approaches both predict amorphous structures based on the potential energy landscape and physical simulation of quenching. RMC simulation, on the other hand, is an optimization method the minimizes the difference between simulated and experimental data to search for the structures that are most consistent with the experimental observations. In most RMC studies (and all those done on a-TiO$_2$) no interatomic potential or other information about potential energy is used, which results in models with potentially unphysical bonding.[29,30] More accurate determination of the amorphous structure can potentially be obtained by integrating experimental data and theoretical constrains. Including multiple experimental data can result in structures that are consistent with all the experimental observations within their uncertainty. Including theoretical constraints can avoid the presence of unphysical structures which nevertheless are consistent with the experimental data. A recent developed open-source structure optimization package StructOpt was designed to enable structure optimization with guidance from multiple experimental data and the potential energy at the same time.[34] The StructOpt package was improved and applied to study the a-TiO$_2$ system in this work.

As mentioned above, understanding the atomic structure of a-TiO$_2$, especially at the MRO length scale, is required to establish important structure-property relationships, which are important to enhance the stability and longevity of the a-TiO$_2$ in PEC. Here we reported a realistic a-TiO$_2$ model structurally consistent with experimental observed SRO and MRO, and which produces a band gap comparable to the experiments when treated with accurate ab initio methods. In this work, the a-TiO$_2$ thin films (~17 nm) were synthesized via the atomic layer deposition. Local structure ordering was characterized based on electron nanodiffraction in the 4-dimensional scanning transmission electron microscopy (4D-STEM) setting, and the MRO structure feature was revealed by the normalized variance $V(k)$ analysis. Angular characterization of the MRO was performed with the angular correlation analysis of nanodiffraction patterns. Incorporating the experimentally determined MRO information from this work as well as SRO PDF information from the literature,[29] an ~8 nm$^3$ (~2 nm on a side) periodic unit cell model of a-TiO$_2$ was optimized by StructOpt. StructOpt attempted to minimize the difference between the simulated and experimental MRO ($V(k)$) and SRO ($G(r)$) while simultaneously minimizing the potential energy. The StructOpt a-TiO$_2$ structure was compared with the melt-quenched a-TiO$_2$ model obtained from IPMD simulation and the crystalline phases, and differences in atomic structure and electronic properties are discussed. The StructOpt a-TiO$_2$ model assures an atomic structure consistent with the potential energy and gives significantly better agreement with MRO compared to the melt-quenched model. Thus, we propose that the StructOpt a-TiO$_2$ model developed here provides a more realistic structure model for amorphous TiO$_2$ film than previously obtained by melt quenching and RMC.

## 2. Experimental Methods

### 2.1 Synthesis of free-standing TiO2 amorphous thin film

Free-standing amorphous TiO$_2$ was prepared via ALD on a thin poly methyl methacrylate (PMMA) sacrificial layer on Si wafer. PMMA layer was fabricated by a spin-coating PMMA solution onto silicon wafer at 3000 rpm with 30 s. The PMMA solution was made by dissolving PMMA powder (molecular weight ~46,000) in toluene (anhydrous 99.5%) with 2wt% concentration. A homemade ALD system was used to deposit amorphous TiO$_2$ film on PMMA film. The PMMA/Si substrate was loaded on a quartz boat and placed at the position 5 cm away from the precursor inlet nozzle. N$_2$ gas with a flow rate of 40 sccm was introduced into the chamber as the carrier gas, which generated a base pressure of 680 mTorr. Note that the growth temperature is a critical factor of the degree of the crystalline within the deposited thin film. In order to obtain a fully amorphous thin film, the chamber temperature was set at 100 °C during the ALD process. Titanium tetrachloride (TiCl$_4$, Sigma-Aldrich, 99.9%)) and DI H$_2$O vapors were exposed in the chamber alternatively with a pulsing time of 0.5 s, respectively, and separated by 60 s N$_2$ purging. Therefore, one ALD cycle involved 0.5 s of H$_2$O pulse + 60 s of N$_2$ purging + 0.5 s of TiCl$_4$ pulse + 60 s of N$_2$ purging and continuously 200 cycles were run in this work. To obtain free-standing TiO$_2$ amorphous films, wafer samples were immersed in chloroform to dissolve PMMA at room temperature. After 12 hours, TiO$_2$ film could be released and readily transferred to SiO$_2$ wafers or TEM grids for characterizations. Atomic force microscopy (AFM) topography was conducted using a XE-70 Park System. The a-TiO$_2$ films on TEM grid were prepared by the same releasing approach for following 4D-STEM characterization. Scanning electron microscopy (SEM) images and energy dispersive X-ray spectroscopy (EDS) performed at 10 kV shows that the synthesized a-TiO$_2$ shows an overall structural and chemical homogeneity over an area of ~10 mm. The SEM image of the as-transferred TiO$_2$ thin film showed a smooth surface with large-area continuity and the corresponding EDS mapping also exhibited the uniform distribution of Ti element in the TiO$_2$ film on SiO$_2$ wafer (Fig. 1a). AFM topography revealed that the film thickness was ~17 nm (Fig. 1b), corresponding to 0.85 nm per cycle, which is consistent to a typical ALD growth rate. Quantitative analysis of the X-ray photoelectron spectroscopy (XPS) spectrum showed that the O/Ti ratio at the surface was about 1.91, which is close to stoichiometry and an indicative of a small amount of oxygen vacancy.

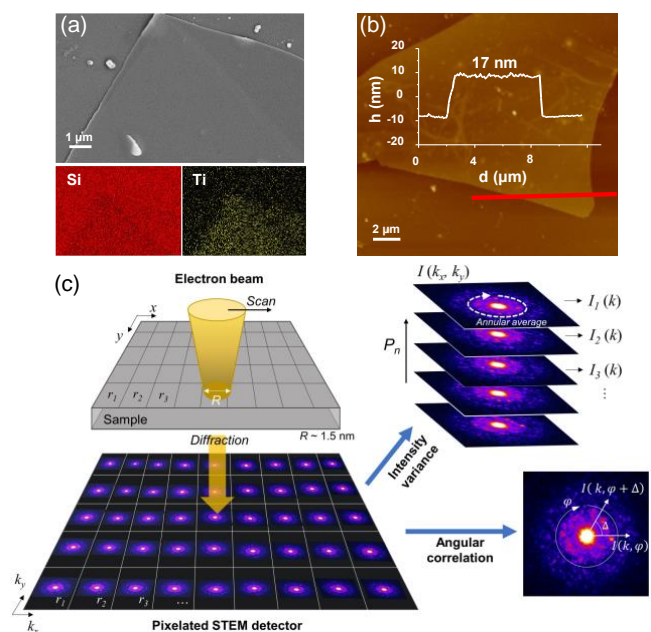

**Figure 1.** (a) SEM image and corresponding EDS elemental mapping of the free-standing amorphous TiO$_2$ film on SiO$_2$ wafer. (b) AFM topography image of free-standing amorphous TiO$_2$ film. Inserted chart is the thickness measurement along the red line. (c) Schematic of 4D-STEM for nanodiffraction acquisition, intensity variance analysis, and angular correlation analysis.

### 2.2 Experimental 4D-STEM measurement of medium-range ordering

4D-STEM records a large number of 2D nanodiffraction patterns in reciprocal space from a 2D real space area of the sample.[26,27,35] We performed 4D-STEM using the Thermofisher Titan STEM operated at 300 kV and equipped with an Electron Microscope Pixel Array Detector (EMPAD)[36] that records scattered electrons with high dynamic range (32 bit) and fast readout speed (1,000 frames per second. As shown in Fig. 1c, approximately 256 by 256 positions were probed within 185 by 185 nm area, and the probe size ($R$) of 1.5 nm was formed using a 10-micrometer condenser aperture. We used a low electron dose (50 pA) and a fast dwell time (4 ms) to eliminate the potential beam damage. This process was repeated over multiple areas to generate the total number of nanodiffraction patterns of ~100,000 per sample. The fast scan and high dynamic range ensure maximizing signal to noise ratio and minimize potential radiation damages to the sample. With the current setup, we observed no visible damages to the sample and no peak shift in V(k) (described below) before and after the beam exposure, which confirms that the radiation damage was effectively prevented and has negligible effect on the MRO measurements. These abundant nanodiffraction patterns were then used to calculate the normalized variance ($V$) of scattered intensity ($I$) as a function of the scattering vector magnitude $k$,[25]

$$V(k,R) = \frac{\langle I^2(r,k,R)\rangle_r}{\langle I(r,k,R)\rangle_r^2} - 1, \qquad \text{(Eq. 1)}$$

where $r$ is the location of the probe in real space, and $<>_r$ represents the spatial averaging over $r$. In other words, $V(k)$ measures the fluctuation among many $I(k)$'s from different areas of the sample, which represents the degree of spatial fluctuation of the structure, at different $k$ values. $k$ is the inverse of the spacing of the planar structure that diffracts (via Bragg condition), and therefore it represents the type of the structural ordering. More quantitatively, $V(k)$ relates to many-body correlation functions, such as 3- or 4-body correlation functions among atoms, which contains more details about the arrangement among a group of atoms (i.e. MRO) as compared to PDF that only shows 2-body correlation function.[37]

The same nanodiffraction data were also used to calculate the average angular correlation function within individual nanodiffraction patterns.[28] The angular correlation function from one nanodiffraction pattern determines the autocorrelation among diffraction intensities as a function of the in-plan azimuthal angle for all $k$, which is given by

$$c_k(\Delta) = \frac{\langle I(k,\varphi)I(k,\varphi+\Delta)\rangle_\varphi - \langle I(k,\varphi)\rangle_\varphi^2}{\langle I(k,\varphi)\rangle_\varphi^2} - 1, \qquad \text{(Eq. 2)}$$

where $\Delta$ is the angular difference between two vectors having the same radius $(k)$ and $\varphi$ is the entire azimuthal angle (Fig. 1c). Many $c_k(\Delta)$'s from the entire area of the sample were then averaged to increase statistical significance of the data.[38] We note that the StructOpt optimization (explained below) used $V(k)$ as the input, while the average $c_k(\Delta)$ was used as an extra measure to compare the output structure from StructOpt to the experimentally characterized structure.

## 3. Computational Methods

### 3.1 StructOpt Optimization

StructOpt package is a multiobjective structure optimization method with guidance from both potential energy derived from molecular modeling and multiple types of experimental data. Three modules implemented in the StructOpt were applied in this work: 1) the LAMMPS module to perform structure relaxation and potential energy evaluation by the Large-scale Atomic/Molecular Massively Parallel Simulator (LAMMPS) Package[39] using interatomic potentials; 2) the FEMSIM module to simulate the $V(k)$ based on the atomic structure and calculate the discrepancy between the simulated and experimental $V(k)$s; and 3) the PDFSIM module to simulate the reduced pair distribution function $G(r)$ from the atomic model and calculate the discrepancy between simulated and experimental $G(r)$s. For a given atomic structure, the objective function ($O$) is defined as a combination of the potential energy $E$, the fitness to experimental $V(k)$, $\chi^2_{FEMSIM}$, and the fitness to the experimental $G(r)$, $\chi^2_{PDFSIM}$,

$$O_{(E,V(k),G(r))} = E + \alpha_{FEMSIM}\chi^2_{FEMSIM} + \alpha_{PDFSIM}\chi^2_{PDFSIM}$$
(Eq. 3)

where $\alpha_{FEMSIM}$ and $\alpha_{PDFSIM}$ are the weighting factors for the FEMSIM and PDFSIM terms, respectively. Weighting factors are introduced to give a desired numerical importance of each term. We apply a two-step optimization approach here, 1) fit almost entirely to G(r) with $\chi^2_{FEMSIM} = 10^3$ and $\chi^2_{PDFSIM} = 10^6$ and ran generations until $\chi^2_{PDFSIM}$ stopped changing significantly; 2) fit predominantly to the V(k) and G(r) with $\chi^2_{FEMSIM} = 10^3$ and $\chi^2_{PDFSIM} = 10^3$ for the remaining generations, allowing the energy to effectively break any degeneracies after the experimental data has been matched. Fitting to G(r) is prioritized since the structure and chemistry of the first nearest neighbor shells, which are captured by G(r), are essential to get correct since those often dominate properties. Then, V(k) was the second most important to get correct since it represented direct experimental structural constraints. Finally, we used the energy from the interatomic potential to remove unphysical arrangements through relaxation and choose the most stable structure between structures otherwise nearly degenerate with respect to the experimentally data. We did not want to weight the energy so strongly that it forced to the system toward the ground state as the amorphous structure is a metastable state by definition.

With a population of 10 input structures, the genetic algorithm performs mutations and crossovers on these atomic structures. Mutations modify a certain portion of the atom position in one structure to create a new structure and crossovers combine two structures by patching them together in certain ways to create new structures to the population. For the updated structure population, the atomic structure is firstly relaxed by LAMMPS module by interatomic potential. Based on the relaxed atomic structure, simulated $V(k)$ and $G(r)$ are calculated by the FEMSIM and PDFSIM module, respectively. Fitness evaluates the goodness of the structure, and a fitness score is calculated for each structure according to the objective function $O_{(E,V(k),G(r))}$. This approach is described in more detail in Ref. 34. Minimizing the discrepancy between experimental and simulated $V(k)$'s aims to generate the MRO in the model that is consistent to the MRO in the real a-TiO$_2$ sample which gave rise to the experimental $V(k)$ signal. Including the discrepancy between experimental and simulated $G(r)$'s aims to confine the structure features related to the SRO in the model. Structures with the best fitness scores are selected for the next interaction of optimization. The structure is iteratively modified until it minimizes the objective function.

The potential energy can be evaluated by various levels of approximation with different atomistic modeling methods, among which the interatomic potential is the optimum approach balancing the computational time and accuracy. In this work, the MA potential[40] is applied to perform the structure relaxation and energy evaluation in the StructOpt optimization. T. Kohler et al[33] have compared the a-TiO$_2$ model obtained from classical MD simulation using MA potential with higher level DFT-based simulations, showing that the MA potential is a reliable potential that could reproduce all relevant structural features of a-TiO$_2$ obtained by quantum-mechanical simulations and experimental $G(r)$.[33] Since the mutations and crossovers can create unphysical structures, it is possible for the initial structures created during the genetic algorithm to have numerical instabilities. Therefore, although we used the MA potential for all final positions and energies, the genetic algorithm created structures were firstly relaxed by a Reax force field potential[41] as this force field remained numerically stable even for quite unphysical structures. The structures were then minimized by the MA potential using conjugate gradient methods with fixed charge on Ti (2.19) and O (-1.098) atoms. Potential energy $E$ in Eq. (3) of the structure was determined by the MA potential.

The simulation of $V(k)$ was performed using the FEMSIM+HRMC code[42,43] using a kinematic diffraction approach.[44,45] The orientations are determined by uniformly sampling the surface of a sphere on a mesh set of taking steps in angles of 4.5 degrees, which ensures the sampling of Bragg diffraction from the model toward all possible directions. 922 unique calculations of different orientations are obtained after accounting all cases. The rotation was to mimic the spatial distribution of the random orientation of MRO in the experimental sample. The diffraction intensity $I(k)$ was calculated from each rotated model via the Dash et. al method[45] by the numeric integration of the local PDF at the position $r$ defined as

$$g_{2A}(r',\boldsymbol{r}) = \sum_j \sum_i A(2\pi Q|\boldsymbol{r} - \boldsymbol{r}_j|)A(2\pi Q|\boldsymbol{r} - \boldsymbol{r}_i|)\delta(|\boldsymbol{r}_i - \boldsymbol{r}_j| - r'),$$ (Eq. 4)

where $A$ is the Airy function (Fourier transform of the aperture function), $r'$ is the real space distance from $\boldsymbol{r}$, and $Q$ is the aperture size that is $Q = 0.61/R$ ($R$ is the probe size, or the spatial resolution). $g_{2A}$ shows PDF among the atoms projected on a 2-dimensional plane, which is the result of the flat Ewald sphere approximation. The variance of the electron diffraction patterns was then calculated among $I(k)$'s for all orientations using Eq. 1. After that, the fitness between the experimental and simulated $V(k)$ was calculated as

$$\chi^2_{FEMSIM} = \min_\beta \left\{ \sum_k^{N_k} \left(\frac{\beta V_{exp}(k) - V_{sim}(k)}{\sigma_k}\right)^2 / N_k \right\},$$ (Eq. 5)

where the summation over $k$ represents the discretized $k$ points at which the experiment was conducted, $\sigma_k^2$ is the experimental error at point $k$, and $\beta$ is a scaling factor proportional to the ratio of experimental to simulated sample thickness $t_{exp}/t_{sim}$. Starting from an initial guess of approximately $\beta_i = 3 * t_{exp}/t_{sim}$,[46] the optimum value of $\beta$ was determined during the StructOpt optimization to minimize the value of $\chi^2_{FEMSIM}$. Evaluating the $\chi^2_{FEMSIM}$ with the scaling factor $\beta$ in the range of $(1 - 0.3)\beta_i < \beta < (1 + 0.3)\beta_i$, the $\beta$ value with the minimum $\chi^2_{FEMSIM}$ is the most suitable scaling factor. The summation is averaged over the total number of $k$ points $N_k$. Some approximations, such as a flat Ewald sphere,[45]

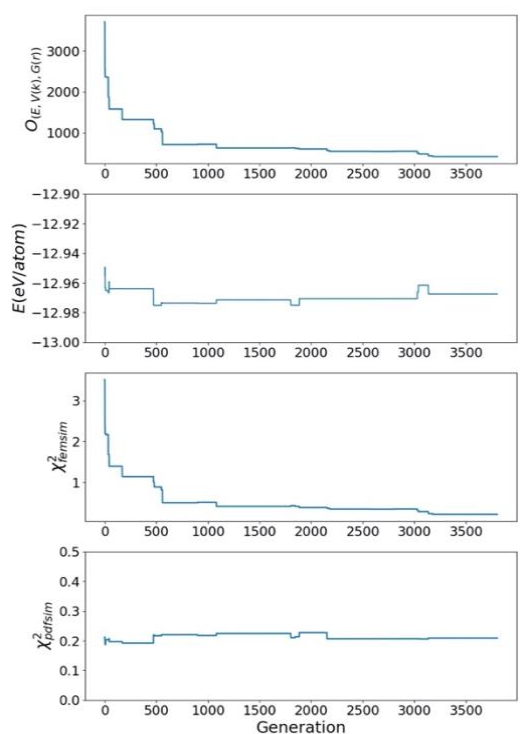

**Figure 2.** The objective function, potential energy, fitness to experimental $V(k)$, and fitness to experimental $G(r)$ of the best scoring structure to that point along the optimization generations.

were applied in the FEMSIM tool, so small errors might be included in the simulated $V(k)$.

The reduced PDF $G(r)$ is calculated by the DebyePDFCalculator within the diffpy.srreal package,[47] where $G(r)$ is calculated as a reciprocal-space Debye summation with a Fourier transform of the total scattering structure function. Hence the $G(r)$ calculation is normalized in the same way as the experimental data to enable a meaningful comparison between simulation and diffraction data.[48] The fitness between the experimental and simulated $G(r)$ is calculated as

$$\chi^2_{PDFSIM} = \min_{a'}\{\sum_i^{N_r}(G_{exp}(r_i) - G_{sim}(r_i, a'))^2/N_r\}, \text{ (Eq. 6)}$$

where the summation over the integer $i$ represents a set of discrete radial points, $r_i$ from 1.65 Å to 7.0 Å with an interval of 0.05 Å. These points were given by the starting point of the first peak and the end of the last peak of G(r) to avoid instrumental noise. The experimental data was reproduced from Petkov, *et al.*, 1998. The minimization is with respect to a bounded cubic cell length, $a'$, which we now explain. Considering the density difference between experimental sample and the modeling, as well as the instrumental error from the experiments, the standard cubic cell size $a$ (described below) used in the rest of the modeling was allowed to vary as part of optimizing $\chi^2_{PDFSIM}$. Specifically, the fitness we minimized with respect to $a'$ in the range $(1-0.2)a < a' < (1+0.2)a$.

In this work, a melt-quenched a-TiO$_2$ model using the "melt-and-quench" method is obtained using the MA potential and serves as one of the input structures for the StructOpt optimization and is also used for comparisons. The 5×5×2 anatase TiO$_2$ model containing 600 atoms was melted in a NVT ensemble at 5000 K for 50 ps with a timestep of 0.5 fs, and cooled down to 3000K with a timestep of 0.5 fs and 200 ps simulation time. Next, the model was equilibrated at 3000 K for another 50 ps. Finally, the model was annealed from 3000 K down to 300 K at a cooling step of 1 K/ps, equilibrated at 300 K for 100 ps, and statically optimized to minimal energy to obtain the final quenched atomic structure. The simulation model used a fixed volume (density) for the V(k) and interatomic potential calculations. This volume was set as a cubic with a supercell edge length of $a$ = 19.05 Å and included 600 atoms. The lattice parameter was determined so as to match a typical experimental density value of a-TiO$_2$ at room temperature (3.84 g/cm$^3$).[49] The model was chosen to be large enough to include typical medium-range ordering length scales and also to keep the computational cost manageable. Simulated $V(k)$ and $G(r)$ were calculated on melt-quenched models with different size from 10 Å to 50 Å and found to be consistent (Fig. S2, S3), showing that there is no significant approximation associated with the calculation of these properties associated with our use of the ~19.05 Å supercell edge length in the model. Structopt optimization starts with the melt-quenched a-TiO$_2$ model and random a-TiO$_2$ structures. Fig. 2 displays the changes of the overall objective function, potential energy, fitness to experimental $V(k)$ ($\chi^2_{FEMSIM}$), and fitness to experimental $G(r)$ ($\chi^2_{PDFSIM}$) of the best scoring structure to that point along the optimization process. The overall objective function decreases until it converges, where the finesses to $V(k)$ and $G(r)$ both reach to its minimum. The optimal structure that best describes the experimental data and is also energetically favorable is obtained by optimizing the objective function following the approach described in Sec. 3.1.

### 3.2 Calculating angular correlation from atomic models with multislice simulation of electron nanodiffraction

In order to compare the symmetry of the MRO generated in the model and compare it to the symmetry observed in experiments, we simulated the nanodiffraction patterns using Multislice simulation and calculated the angular correlation function, $c_k(\Delta)$, from those nanodiffraction patterns. Unlike in the diffraction intensity simulation in FEMSIM where a flat Ewald sphere is assumed, Multislice simulation generates nanodiffraction patterns with fully realistic conditions, including dynamic diffraction.[50] It treats the atomic model as a collection of atomic layers (slices) perpendicular to the direction of incoming electrons, generating the scattered electron waves from each slice based on weak phase approximation, consecutively through the entire stack of layers, which can take both multiple and thermal scattering into account. The simulation was used to generate the nanodiffraction patterns from the atomic structural models generated by StructOpt optimization as well as the MD simulation with the atomic potential only. For the a-TiO$_2$ models which has the cubic size of 19.05 Å, the layer thickness was set to be less than 2 Å. The same model rotation scheme that we used in FEMSIM (see Sec. 3.1) is applied in multislice simulation. The overall amplitudes of the experimental and simulated $c_k(\Delta)$'s should be different because the size of the model used in the simulation is smaller than the thickness of the experimental sample. To accommodate this difference, the simulated $c_k(\Delta)$ was scaled by a factor of 10, and a qualitative comparison between the experiment and simulation was made, which will be shown in Section 4.3.

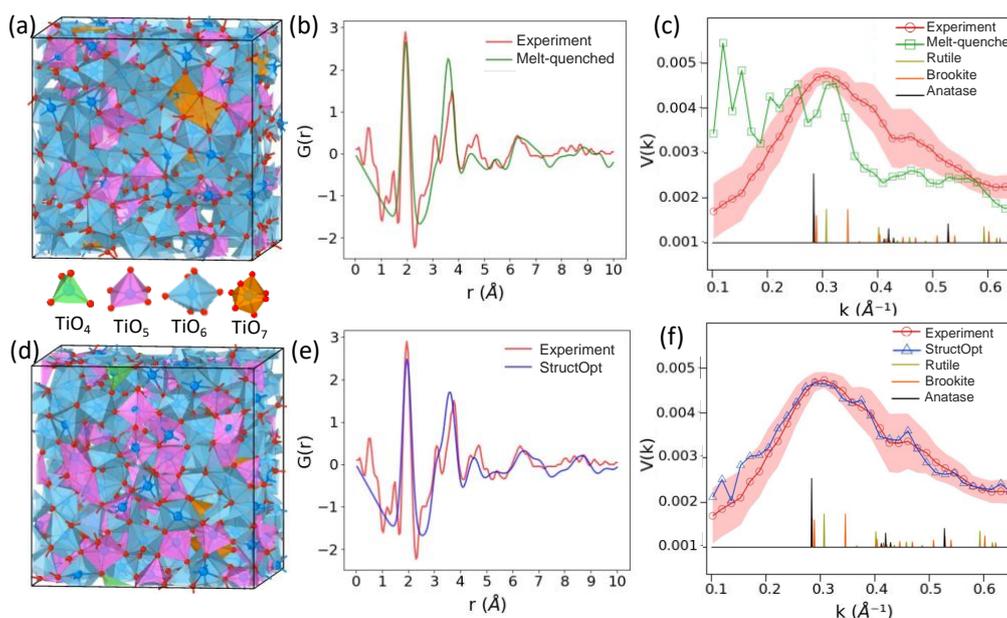

**Figure 3.** Polyhedral visualization of (a) the melt-quenched a-TiO$_2$ structure and (d) the StructOpt a-TiO$_2$ structure, respectively. Comparison between the experimental G(r) with (b) simulated G(r) of the melt-quenched a-TiO$_2$ model and (e) simulated G(r) of the StructOpt a-TiO$_2$ structure, respectively. Comparison of the experimental V(k) to (c) simulated V(k) of melt-quenched a-TiO$_2$ model and (f) simulated V(k) of the StructOpt a-TiO$_2$ model, respectively. Red shade represents the uncertainty range of the experimental V(k). Scaled XRD patterns of Rutile, Anatase, and Brookite phases are shown as vertical lines. G(r) was reproduced from Ref. 29.

### 3.3 DFT calculation of electronic properties

Electronic properties of the StrucOpt a-TiO$_2$, melt-quenched a-TiO$_2$ model, and the crystalline phases were studied using density functional theory (DFT) implemented by the Vienna ab initio simulation (VASP) package.[51] The Heyd–Scuseria–Ernzerhof (HSE06) hybrid functional[52] with 25% Hartree-Fock exchange was applied to calculate the electronic structure since the band gap is commonly underestimated in LDA and GGA. The valence electron configuration of the Ti and O atoms utilized in all calculations were $3p^64s^23d^2$ and $2s^22p^4$, respectively. The cutoff energy of the plane wave is 400 eV, and the convergence of the electronic self-consistent energy is $10^{-4}$ eV. Calculations were performed on the Γ point in the k-space and all the calculations were spin polarized.

### 3.4 Integration of modeling and experiments

This work involves extensive use of or comparison to experimental data, including density, $G(r)$, $V(k)$, angular correlation functions, and band gap. Most experimental data used in this paper comes from our own studies of a 17 nm thin film a-TiO$_2$ to provide a consistent set of data associated with the particularly amorphous structure. However, we also make use of density, $G(r)$, and band gaps from previous experiments on other a-TiO$_2$ samples. We believe that these quantities are relatively insensitive to the exact processing that led to the a-TiO$_2$ sample and therefore similar enough across samples that the use of data from different studies is a reasonable approximation. One exception might be density, which ranges from 3.02 to 3.92 g/cm$^3$,[49,53] but this value is only used to set our simulation unit cell volume and our structures and their properties are not highly sensitive to this value.

## 4. Results and Discussion

### 4.1 Optimization results

Fig. 3d present the structure obtained from the StructOpt optimization to minimize energy and maximize agreement with experimental $G(r)$ and $V(k)$ (see Sec. 2.2), referred as the StructOpt a-TiO$_2$ model. Optimization of the $V(k)$ is expected to confine the structure with more realistic MRO compared to the previous RMC simulations[29,30] which were only based on PDF. The melt-quenched model obtained by the MA potential is analyzed as a reference shown in Fig. 3a. The results show that both the melt-quenched and the StructOpt a-TiO$_2$ model consist of distorted TiO$_X$ structural units (Fig. 3a, 3d), and agree reasonably well with the experimental $G(r)$ data (Fig. 3b, 3e). However, the melt-quenched model shows significant discrepancy with respect to the experimental $V(k)$ ($\chi^2_{FEMSIM} = $ 5.14, Fig. 3c), indicating that the use of the potential and fast quenching only does not generate MRO structure consistent to the MRO signal from the experimental a-TiO$_2$. With optimizing the experimental $V(k)$ data during the StructOpt optimization, the StructOpt a-TiO$_2$ model shows overall good agreement to the experimental $V(k)$ ($\chi^2_{FEMSIM} = 0.22$, Fig. 3f).

**Table 1.** Comparison between the StructOpt a-TiO$_2$ model, the melt-quenched a-TiO$_2$ model, and the crystalline phases for nearest Ti-O pair distance, fraction of polyhedra, and the averaged shared vertex/edge between polyhedra.

| | Ti-O distance | CN | TiO$_4$ | TiO$_5$ | TiO$_6$ | TiO$_7$ | Shared vertex | Shared edge |
|---|---|---|---|---|---|---|---|---|
| StructOpt a-TiO$_2$ | 1.95 | 4.67 | 0.5% | 28% | 68.5% | 2.5% | 5.64 | 3.86 |
| melt-quenched a-TiO$_2$ | 1.95 | 4.67 | 0 | 22% | 74.5% | 3.5% | 5.77 | 3.83 |
| Anatase | 1.95 | 6 | 0 | 0 | 100% | 0 | 4 | 4 |
| Rutile | 1.96 | 6 | 0 | 0 | 100% | 0 | 8 | 2 |
| Brookite | 1.96 | 6 | 0 | 0 | 100% | 0 | 6 | 3 |
| Exp.[29] | 1.89 | 4.5±0.4 | | | | | | |

### 4.2 Structure analysis

To understand the structure difference arisen from the melt-quenched to the StructOpt a-TiO$_2$ model, a comparison of the structural features between the amorphous and crystalline TiO$_2$ are summarized in Table 1. Firstly, the nearest Ti-O bonding distance is populated around 1.95 Å for both the StructOpt and melt-quenched a-TiO$_2$ model, which is close to the first Ti-O bond distance in a crystalline Ti-O phase, indicating that the a-TiO$_2$ has similar short-range ordering with the crystalline phase. Average coordination number (CN) of Ti is calculated by integrating the area underneath the first peak from the simulated PDF. Both the melt-quenched and the StructOpt a-TiO$_2$ model show consistent results with the previous experiments.[29,54] Though the melt-quenched and StructOpt model show very close structure features at the short-range scale, there is a clear difference between the building blocks in terms of the fraction of different types of TiO$_X$ polyhedra. Compared with the melt-quenched model, there is a significant increase of 5-coordinate polyhedra and decrease of 6-coordinate polyhedra (octahedra) in the StructOpt model. Octahedra are the building blocks of TiO$_2$ crystalline phases. Since the atomic potentials are largely developed based on the crystalline phases,[40] it is not surprising that the melt-quenched model, which only relies on the interatomic potential, forms high population of octahedra. However, the population of octahedra in the StructOpt model decreases after including the experimental MRO features.

The average numbers of shared vertex and edge between the TiO$_X$ units are counted as indicators of polyhedral connectivity.[32] Average shared vertex and edge of the StructOpt a-TiO$_2$ model is 5.64 and 3.86, respectively. These results are quite similar to the melt-quenched structure. Compared to the crystal phases, the StructOpt a-TiO$_2$ shows a structural similarity to the brookite phase, which has a shared vertex of 6 and shared edge of 3. The structural similarity is also observed in the bond-angle distribution function shown in Fig. 4. The Ti-O-Ti angle describes the connectivity between TiO$_X$ units in the system. In general, Ti-O-Ti angle has two distinct peaks corresponding to two common types of the linking between TiO$_X$ units via edge and vertex. For the StructOpt model, the first peak is populated around 100°, and the second peak populates from to 118° to 136°. Distribution of these two major peaks match well with the brookite phase, indicating the StructOpt model has a similar TiO$_X$ connection network with the brookite phase.

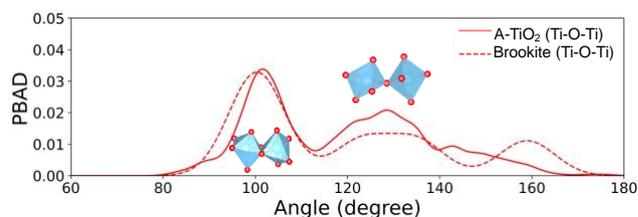

**Figure 4.** Normalized density of the Ti-O-Ti bond-angle distribution function of the StructOpt amorphous TiO$_2$ model compared with Brookite phase. Gaussian kernel-density estimation[55,56] is applied for the distribution of Ti-O-Ti bond-angle. Inserted graphs present two TiO$_6$ connected via sharing edge and vertex.

Next, we discuss the peak positions in the $V(k)$ data in Fig. 3c and 3f, and their implications to the MRO structure. In the experimental data shown in Fig. 3c (red curve), two broad peaks in $V(k)$ are observed, one at ~ 0.3 Å$^{-1}$ and the other small peak at ~ 0.46 Å$^{-1}$. The magnitude of each peak is related to the degree of structural fluctuation created by the distribution of MRO domains, and the position of the peak at certain $k$ relates to the type of the MRO, since $k = 1/d$ where $d$ is the real space spacing between the planes within the ordering. The fact that the experimental variance peaks are at ~ 0.3 and ~ 0.46 Å$^{-1}$ suggests that those are the $k$ positions that relate to the internal structure (and therefore the type) of the MRO. In many amorphous systems, the structure of measured MRO approximately resembles that of crystalline phases,[57] which has also been suggested to be the case in amorphous TiO$_2$. Based on this assumption, we compared the XRD peak positions of the known TiO$_2$ crystalline phases[63–65] to the $V(k)$ data, and all are shown in Fig. 3c. The signature peak position of experimental $V(k)$ at ~ 0.3 approximately match with the strongest known anatase, brookite, and rutile peak positions, which suggests that the structure of MRO in TiO$_2$ resembles that of those known phases. However, the peaks in the $V(k)$ are very broad, indicating that the MRO structure is substantially disordered, despite its resemblance to the crystalline phases. Meanwhile, the two amorphous models – melt-quenched and StructOpt model, show significant differences in terms of how their simulated $V(k)$ fits to the experimental $V(k)$. The simulated $V(k)$ from the melt-quenched model shows poor fit to experimental $V(k)$ (Fig. 3c), suggesting that the MRO structure of the model is not consistent to that of the real sample. In particular, the simulated $V(k)$ has the major peak positions at low $k$ range, about 0.1 – 0.2 Å$^{-1}$, where there is no crystalline diffraction peak occur. This implies that the structural origin of the peaks at low $k$ are not based on the MRO that resembles crystalline ordering. Instead, it is possible that the low $k$ peaks might have been generated by some voids within the model.[66] On the other hand, the StructOpt model shows simulated $V(k)$ that matches the experimental $V(k)$, with the major peak positions matching the positions of crystalline diffraction peaks, suggesting that the dominant structural features in the model structure that gives rise to peaks in $V(k)$ are the crystal-like MRO.

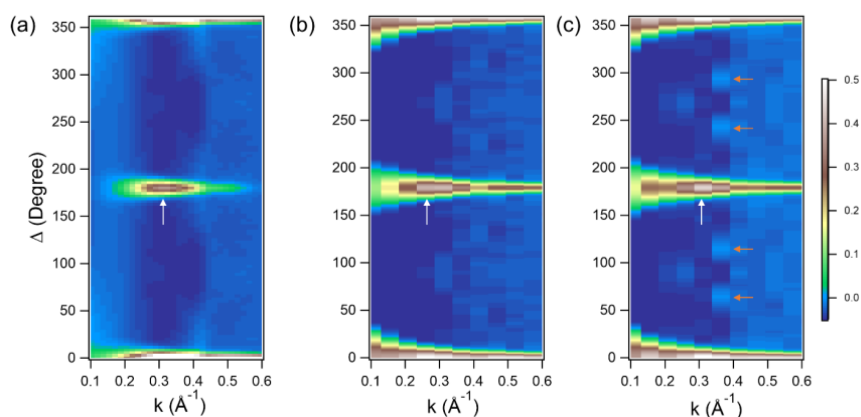

**Figure 5.** Averaged angular correlation plots for (a) experimental a-TiO$_2$ film, (b) the melt-quenched model, (c) the StructOpt model.

### 4.3 Angular correlation analysis of nanodiffraction patterns

Angular correlation, $c_k(\Delta)$ (Eq. 2) from both the experimental and simulated diffraction patterns were calculated. The experimental $c_k(\Delta)$ (Fig. 5a) shows a peak at 180° location, which indicates 2-fold symmetry at $k$ ~ 0.31 Å$^{-1}$ (white arrow). The $k$ position is consistent with the peak position in experimental $V(k)$ shown in Fig. 3c, which indicates that the MRO that gives rise in $V(k)$ has 2-fold rotational symmetry (in other words, it has a plane-like MRO that diffracts the electron beam). Meanwhile the $c_k(\Delta)$'s calculated from the multislice simulated diffraction patterns from both the melt-quenched model (Fig. 5b) and StructOpt model (Fig. 5c) show strong 2-fold symmetry as well, consistent to the experimental data. However, $c_k(\Delta)$ from the melt-quenched model has the peak at $k$ ~ 0.26 Å$^{-1}$ (Fig. 5b), which is slightly lower than the peak position in the experimental data. This discrepancy is corrected in the StructOpt model (Fig. 5c) where the peak position in the $c_k(\Delta)$ matches the experimental data at $k$ ~ 0.31 Å$^{-1}$. The result here suggests that in finding a better fit to $V(k)$ the StructOpt optimization also improved the match to the experimental angular correlation. Interestingly, the $c_k(\Delta)$ from the StructOpt model (Fig. 5c) also shows a few brighter spots at higher $k$ (~ 0.37 Å$^{-1}$, orange arrows), with an apparent 6-fold symmetry, that are not present in the melt-quenched $c_k(\Delta)$. This result indicates that when StructOpt optimizes the model to fit $V(k)$ (even though $V(k)$ only has one-dimensional information, as shown in Fig. 3c), the combination of the atomic potential and $V(k)$ during the optimization created a higher-order symmetry in the MRO structure. Such a higher order symmetry, however, is not observed in the experimental data (Fig. 5a), possibly because the experimental data were averaged over a much larger volume of the material as compared to the simulated volume of the material shown in Fig. 5c. In other words, although Fig. 5c may be showing a possible MRO symmetry, it may not be significant in terms of volume fraction in the real sample to appear in the averaged experimental pattern shown in Fig. 5a. We note that these spots are quite weak, with a local peak intensity much lower than the peak intensity associated with the 2-fold symmetry, which may make them difficult to resolve experimentally and easily generated by a very modest level of correlation in the StructOpt model.

### 4.4 Electronic structure

To further explore the quality of the StructOpt a-TiO$_2$ structure, we studied its electronic structure and compared it to that from the crystalline phases, melt-quenched a-TiO$_2$ structure, as well as the experimental results. As shown in Table 2, the calculated band gaps for the crystalline phases in this work agree well with the experimental results. The amorphous models show slightly smaller band gaps compared to the crystalline phases, which was expected due to the undercoordination Ti atoms and the distortion of the polyhedral.[67–70] Similar with the crystalline phases, the valence band maximum (VBM) and the conduction band minimum (CBM) of the StructOpt a-TiO$_2$ model mainly consist of O-2p and Ti-3d orbitals, respectively (Fig. 6). These localized electron levels near the VBM and CBM created tails that extended the valence and conduction band and reduced the electronic band gap. Defect states arise from the undercoordinated Ti atoms might locate near the band edge,[71] accordingly the simulated band gap of a-TiO$_2$ is slightly smaller compared to the experimental results.

The StructOpt a-TiO$_2$ model provides detailed geometric and electronic structure information within a unit cell of 600 atoms, which is much larger than the previous models predicted by ab initio simulations. In addition, it is expected to be more accurate compared to the previous models resulting from melt-quenched AIMD and IPMD simulation and RMC simulation. This StructOpt a-TiO$_2$ model presents good agreement with the experimental observations including the SRO, MRO ($V(k)$ and angular correlations. Nevertheless, the model does have some remaining issues. First, the 600-atom model is still smaller than ideal for capturing the experimental measured MRO, which was acquired over a large area of the a-TiO$_2$ film, and larger cells should be explored in the future. Second, the angular correlation of the model shows a minor 6-fold symmetry that does not present in the experimental V(k) and this should be investigated to determine if it a sign of structural mismatch between the model and experiment or more an issue of the type of averaging that occurs in the experiment vs. the model. In future, a multislice simulation of the angular correlation function could be included into the StructOpt optimization process to allow more accurate fitting to experiments. Finally, it is almost certain that that this model is not unique and that the true system has significant variations in different nanoscale regions. This model therefore represents at best a sampling of a typical local environment from the true a-TiO$_2$ structure. The model may therefore not be suitable for studying properties where the heterogeneity of local environments is important, such as dynamic properties (e.g., atomic diffusion), shear band nucleation and evolution, or nucleation and growth during crystallization. However, the bond length, coordination number, many aspects of electronic structure, etc. which are not too sensitive to the model size are expected to be well described by this model. In particular, the band gap/band edge position, which are believed most relevant to a-TiO$_2$'s charge transport properties and its stability in the PEC process, are likely to be well reproduced by having accurate local atomic environments. This property will allow the model to be useful for further studies in the PEC field. Overall, we believe that the present model is the best structural model we have for a-TiO$_2$ to date and can help with future studies of the structure-property relationships.

**Table 2.** Calculated band gap of Anatase, Brookite, and Rutile phases, and the StructOpt amorphous TiO$_2$ compared with experimental results.

| | Anatase | Brookite | Rutile | Melt-quenched a-TiO$_2$ | StructOpt a-TiO$_2$ |
|---|---|---|---|---|---|
| HSE | 3.17 eV | 3.51 eV | 3.06 eV | 2.98 eV | 2.74 eV |
| Experiment | 3.20 eV [58,59] | 3.00-3.40 eV [59] | 3.00 eV [59,60] | 3.18-3.34 eV [8,61,62] | |

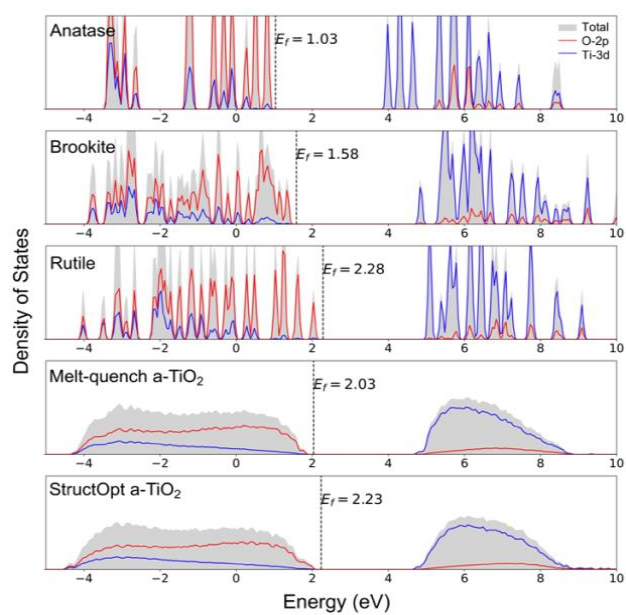

**Figure 6.** Density of states of the Anatase, Brookite, Rutile, melt-quenched, and StructOpt a-TiO$_2$ models calculated by DFT at with HSE06 hybrid functional. Energy level was aligned to the O 2s core level, and the Fermi energy level (E$_f$) are labeled.

## Conclusion

An improved atomic model of a-TiO$_2$ compared to previous approaches was obtained by the StructOpt optimization method driven by the experimental data and potential energy constraints. This ~8nm$^3$ cubic model shows structure consistency with experiment observations of short-range ordering and medium-range ordering. To develop this model, firstly, a-TiO$_2$ thin films (~17 nm) were synthesized via the atomic layer deposition and characterized by the 4D-STEM technique. The medium-range ordering (MRO) with planner spacing around ~ 3 Å in the thin film was revealed by the normalized variance $V(k)$ analysis. The angular correlation analysis showed that the MRO has 2-fold symmetry. Then, using the experimental $V(k)$, reduced pair distribution function $G(r)$ (from previous work) as inputs, the a-TiO$_2$ structure was optimized driven by eliminating the difference between the simulated and experimental $V(k)$s, $G(r)$s, as well as minimizing the potential energy evaluated by an interatomic potential. The StructOpt a-TiO$_2$ model was compared with a melt-quenched a-TiO$_2$ obtained by simulation with interatomic potential only, and the crystalline phases. The melt-quenched a-TiO$_2$ model agrees with the $G(r)$ but does not match the $V(k)$. While the StructOpt a-TiO$_2$ model displays agreement with both the experimental $V(k)$ and $G(r)$. The structural network analysis indicates that the StructOpt a-TiO$_2$ has a structural similarity to the Brookite phase. Finally, band gap is investigated by DFT calculation with HSE06 hybrid functional. The calculated band gap of the StructOpt a-TiO$_2$ model was 2.74 eV, slightly smaller than the experimental measurements and crystalline phases which is likely due to the localization of the energy levels near the band edge. The StructOpt a-TiO$_2$ model provides realistic local structure information consistent with experiments, and therefore should be useful for further studies about a-TiO$_2$ structure-property relationships.

## Data and Codes Information

The raw and processed data required to reproduce these findings are available to download from figshare at https://figshare.com/account/home#/projects/136900
The StructOpt code is available at GitHub https://github.com/mengjun930/StructOpt_modular

## Corresponding Author

*Dane Morgan: ddmorgan@wisc.edu
*Jinwoo Hwang: hwang.458@osu.edu
*Jun Meng: jmeng43@wisc.edu

## Author Contributions

J. M. and M. A. contributed equally to this work. X. W., J. H., and D. M. designed and conceptualized this project. Y. D. synthesized the amorphous TiO$_2$ thin film and investigated the structure morphology using the SEM, EDS, AFM, and XPS. C. C helped with the thickness measurement of TiO$_2$ thin film with AFM. M. A. investigated the a-TiO$_2$ film with 4D-STEM, and analysed the results. J. M. did the structure optimization, electronic property calculations of a-TiO$_2$, and analysed the results. J. M. and M. A. wrote the manuscript with inputs from all authors. D. M and J. H. supervised the research, reviewed, and edited the manuscript.

## Conflicts of interest

There are no conflicts to declare.

## Acknowledgment

We gratefully acknowledge funding from the U.S. Department of Energy (DOE), Office of Science, Basic Energy Sciences (BES), under Award # DE-SC0020283. This research was also performed using the computer resources and assistance of the UW- Madison Center for High Throughput Computing (CHTC) in the Department of Computer Sciences. Electron microscopy was performed at the Center for Electron Microscopy and Analysis at the Ohio State University.

# Experimentally informed structure optimization of amorphous TiO2 films grown by atomic layer deposition

*Jun Meng[1†], Mehrdad Abbasi[2†], Yutao Dong[1], Corey Carlos[1], Xudong Wang[1], Jinwoo Hwang[2\*], Dane Morgan[1\*]*

[1]Department of Materials Science and Engineering, University of Wisconsin-Madison, Madison, WI 53706, United States

[2]Department of Materials Science and Engineering, The Ohio State University, Columbus, Ohio 43210, United States

## XPS spectrums of free-standing TiO$_2$ film

Ti 2p core spectra was fitted with two peaks located at 458.4 eV and 464.2 eV corresponding to Ti 2p$_{3/2}$ and Ti 2p$_{1/2}$ with 5.8 eV peak separation which implied +4 valence for Ti. O 1s spectra showed the peak at 529.9 eV regarded as Ti-O bonds in TiO$_2$. Through the surface quantitative analysis on the integration of fitting peaks divided by the relative sensitive factor, the O/Ti ratio value was about 1.91, indicating stoichiometry TiO$_2$ with tiny oxygen vacancy.

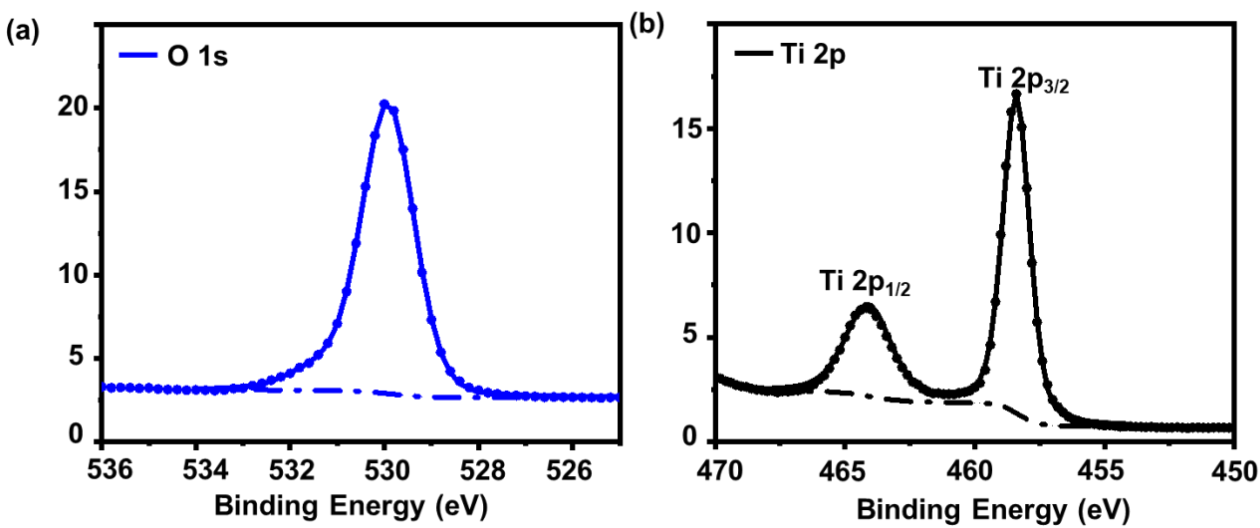

**Figure S1.** XPS spectrums of free-standing TiO$_2$ film grown under 100°C. a, O 1s core spectra. b, Ti 2p core spectra.

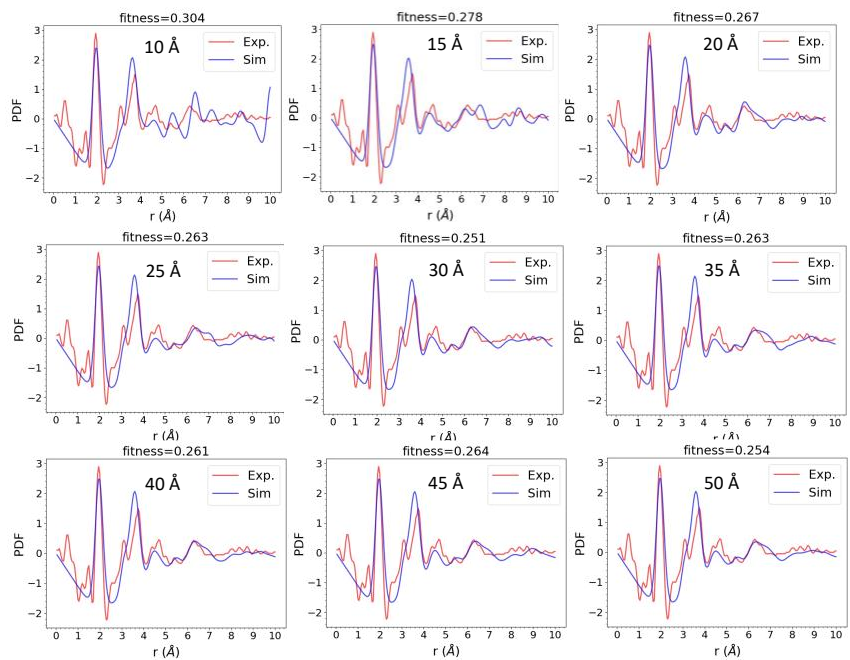

**Figure S2.** Simulated G(r) of melt-quenched a-$TiO_2$ model with different size.

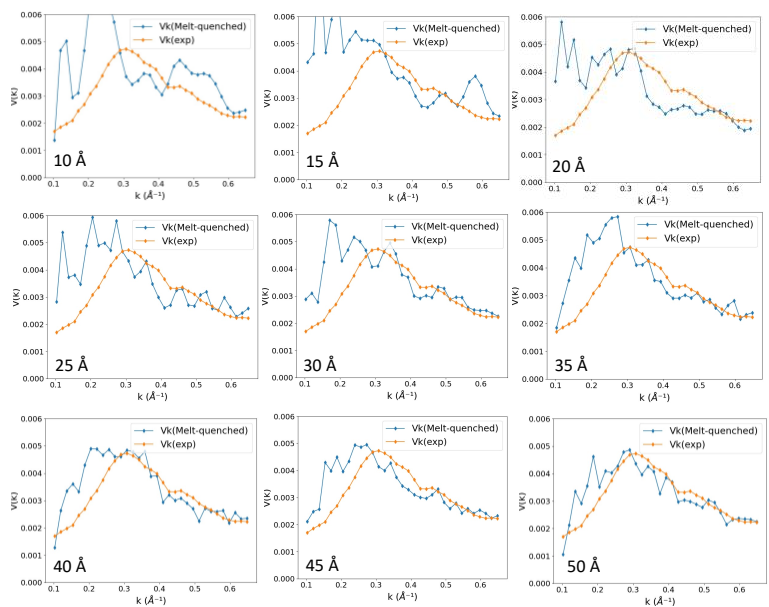

**Figure S3.** Simulated V(k) of melt-quenched a-$TiO_2$ model with different size.